\documentclass[]{Liebert_Author}
\usepackage{amsmath,amssymb,xspace,subfigure}
\usepackage{mathrsfs}
\usepackage{amsfonts}
\usepackage[usenames,dvipsnames]{xcolor}
\usepackage[]{graphicx}
\usepackage{url}
\usepackage{hyperref}
\usepackage[margin=1in,headheight=13.6pt]{geometry}\usepackage{url}
\usepackage{float}
\usepackage{bm}
\usepackage{soul}
\usepackage{setspace}
\setcitestyle{square,comma,numbers}

\graphicspath{{./Figures/}}
\title{Simplified discrete model for axisymmetric dielectric elastomer membranes with robotic applications}
\author{Zhaowei Liu,$^{1}$ Mingchao Liu,$^{2\ast}$ K. Jimmy Hsia,$^{3,4}$ Xiaonan Huang,$^{5}$ Weicheng Huang$^{6\ast}$\\
$^{1}$Department of Engineering Mechanics, Hohai University, \\Nanjing 211000, China\\
$^{2}$Department of Mechanical Engineering, University of Birmingham, \\Birmingham, B15 2TT, UK\\
$^{3}$School of Mechanical and Aerospace Engineering, Nanyang Technological University, \\ Singapore 639798, Republic of Singapore\\
$^{4}$School of Chemistry, Chemical Engineering and Biotechnology, \\ Nanyang Technological University, Singapore 639798, Republic of Singapore\\
$^{5}$Robotics Department, University of Michigan, Ann Arbor\\
$^{6}$School of Engineering, Newcastle University, Stephenson Building, \\ Newcastle upon Tyne NE1 7RU, UK\\
$^\ast$To whom correspondence should be addressed;\\
E-mail: m.liu.2@bham.ac.uk (M.L.) and weicheng.huang@newcastle.ac.uk (W.H.)
}

\begin{document} 
{\setstretch{1.2}
\maketitle

\begin{abstract}
Soft robots utilizing inflatable dielectric membranes can realize intricate functionalities through the application of non-mechanical fields.
However, given the current limitations in simulations, including low computational efficiency and difficulty in dealing with complex external interactions, the design and control of such soft robots often require trial and error.
Thus, a novel one-dimensional (1D) discrete differential geometry (DDG)-based numerical model is developed for analyzing the highly nonlinear mechanics in axisymmetric inflatable dielectric membranes.
The model captures the intricate dynamics of these membranes under both inflationary pressure and electrical stimulation. Comprehensive validations using hyperelastic benchmarks demonstrate the model's accuracy and reliability. 
Additionally, the focus on the electro-mechanical coupling elucidates critical insights into the membrane's behavior under varying internal pressures and electrical loads. 
The research further translates these findings into innovative soft robotic applications, including a spherical soft actuator, a soft circular fluid pump, and a soft toroidal gripper, where the snap-through of electroelastic membrane plays a crucial role. Our analyses reveal that the functional ranges of soft robots are amplified by the snap-through of an electroelastic membrane upon electrical stimuli.
This study underscores the potential of DDG-based simulations to advance the understanding of the nonlinear mechanics of electroelastic membranes and guide the design of electroelastic actuators in soft robotics applications.
\end{abstract}
}

\section{Introduction}
\label{intro}

Electroelastic membranes, made from dielectric elastomers, are a class of advanced materials that possess both mechanical and electroactive properties. These membranes can deform either under  an electric field due to their inherent dielectric properties under mechanical loading due to their elastic properties or under a combination of both. This unique combination allows them to undergo substantial deformation in response to applied voltage changes~\cite{tang2023review}.

In soft robotics, flexible membranes commonly find application in the form of inflatable structures, which can be enlarged or reduced by exchanging fluids such as air or liquid, and frequently experience substantial deformation and potential instabilities~\cite{adkins1952large,hart1967large,yang1970axisymmetrical,pamulaparthi2019instabilities}. In the context of electroelastic membranes, the application of an electric potential difference across the electrodes coated on both sides may result in a notable volume change within the encapsulated fluid. This extraordinary mechanical response to electrical load in electroelastic membranes positions them at the forefront of cutting-edge developments in the application of flexible intelligent structures, such as artificial muscles~\cite{anderson2012multi,qiu2019dielectric} and soft robotics~\cite{tang2023review,majidi2014soft, yang2023morphing} like, e.g., soft grippers~\cite{shian2015dielectric,pourazadi2019investigation, coulson2022versatile} and soft actuators~\cite{zhao2010theory,hau2018novel}. Researchers utilized experimental, analytical, and numerical methods to reveal the sophisticated mechanical behaviors of electroelastic membranes. Goulbourne et al.\cite{goulbourne2005nonlinear} formulated a mathematical framework that integrates substantial deformations, material nonlinearity, and electrical effects. This framework, established by applying Maxwell-Faraday electrostatics and nonlinear elasticity, allows for the analysis of large deformation of electroelastic membranes\cite{goulbourne2007electro}. Fox and Goulbourne~\cite{fox2008dynamic} explored the dynamic response of dielectric elastomer membranes, specifically focusing on  electromechanical behaviors under time-varying voltage and pressure inputs.

Due to the thin and flexible nature, electroelastic membranes inherit the instabilities from hyperelastic membranes. The coupling  between electrical and mechanical responses adds complexity to  stability analysis. Consequently, numerous studies have been undertaken to analyze the instabilities in electroelastic membranes. An experimental study of the behaviors of electroelastic membranes  by Kollosche et al.\cite{kollosche2012complex} illustrates the interplay between limit point and wrinkling instabilities, attributed to the coupling effects. Zhao and Suo\cite{zhao2007method} analyzed the instability of dielectric elastomers, providing insights to inform the design of actuators. Rudykh et al.\cite{rudykh2012snap} introduced a methodology utilizing the snap-through instability of thick-walled electroelastic balloons for the design of actuators. Li et al.\cite{li2013giant} explored significant voltage-induced deformations in dielectric elastomers. Melnikov and Ogden~\cite{melnikov2018bifurcation} introduced a mathematical approach for studying the deformation bifurcation of a  thick-walled cylindrical tube. A comprehensive review of different instabilities of soft dielectrics can be found in~\cite{dorfmann2019instabilities}

To achieve a specific functionality in soft robots and soft actuators,  dielectric elastomer membranes with axisymmetry, e.g., cylindrical~\cite{khayat1992inflation,guo2001large,pamplona2006finite}, spherical~\cite{feng1973contact,akkas1978dynamic,verron1999dynamic}, or toroidal~\cite{pamulaparthi2019instabilities,sanders1963toroidal,liepins1965free,kydoniefs1967finite,li1995finite}, are commonly used as a major structural component. Considering their axisymmetric characteristic, the analysis can be simplified from a three-dimensional (3D) to a two-dimensional (2D) configuration, and a group of ordinary differential equations (ODEs) can be formulated to study its mechanical response; e.g., mechanics of dielectric elastomer membranes \cite{li2013giant}, fluid pump \cite{li2017robust}, and cell cytokinesis \cite{turlier2014furrow}. However, most  aforementioned studies have mainly focused on simple configurations and loading conditions, lacking systematic and comprehensive research on key issues such as geometrically nonlinear dynamics,  nonlinear boundary contact, multi-physical actuation, and mechanical instability of the dielectric elastomer membranes. These considerations are essential for the applications in soft robotics  while almost impossible to incorporate within the traditional ODEs. Moreover,  real-time simulations of  dielectric elastomer actuators are  a prerequisite for the optimal design and online control of these soft machines, and the Finite Element Method (FEM)-based framework may not be computationally efficient in large-scale numerical analysis.

In this study, we introduce a novel 1D discrete model based on the discrete differential geometry (DDG) method for the nonlinear dynamic analysis of axisymmetric dielectric elastomer membranes.This model has applications in soft robotics and soft actuators. The DDG-based simulations are well-suited for handling complex loading conditions, e.g., multi-physical actuation and nonlinear boundary contact, which are important in robotics applications. For example, the DDG-based beam model has been successfully developed to simulate the dynamic motion of shape memory alloy (SMA)-powered soft robots \cite{goldberg2019planar, huang2020dynamic, huang2021modeling} and magnetic-powered soft robots \cite{huang2023discrete, huang2023modeling}. The focus is now turned to the modeling of axisymmetric dielectric membranes and the associated applications in soft robotics.

Following the assumptions introduced in the previous work on the 1D discrete elastic shell model \cite{huang2024discrete}, we use a rotational symmetric curve to describe the axisymmetric membrane, and employ a general nonlinear hyperelastic constitutive law to describe the material's nonlinearity. The electro-mechanical coupling is also included within the discrete model. The gradient and Hession of the elastic and electric potentials are derived analytically, resulting in a fully implicit numerical framework. Next, to demonstrate the accuracy of our developed discrete model, several examples from both theoretical solutions and 3D FEM simulations are adopted for cross-validation. Moving forward, three examples are selected to demonstrate the usability of our model in robotic applications: (i) a circular fluid pump, (ii) a toroidal gripper, and (iii) a spherical actuator. In particular, this simplified 1D model presented here can run faster than real-time on one thread of a desktop processor, which makes it ideally suited for algorithms that iterate over a wide variety of parameters to select an actuator with an optimized design; it can also provide timely feedback for the online control of robotic locomotion.

This manuscript is organized as follows. Section~\ref{sec:Modeldevelopment} illustrates the fundamental aspects of the proposed 1D DDG-based numerical model. The validation by comparing with benchmarks are presented in Section~\ref{sec:Modelvalidation}. Subsequently, Section~\ref{sec:Roboticapplication} demonstrates the real engineering robotic application simulated using the developed numerical model. Finally, the key insights and findings are discussed and concluded in Section~\ref{sec:conclusions}.

\section{Model development}\label{sec:Modeldevelopment}
In this section, we present a procedure for a discrete numerical model based on the DDG framework \cite{grinspun2006discrete} to simulate the inflation of axisymmetric dielectric elastomer membranes, incorporating both pneumatic and electrical actuation and addressing contact challenges.
Adhering to conventional mathematical notation, the square brackets $[ , ]$ are used to group algebraic expressions. Round brackets $( ,)$ are used to denote the dependencies of a function. If brackets are used to denote an interval, then $(,)$ stands for an open interval and $[,]$ is a closed interval. Curly brackets ${,}$ are used to define sets. Also, a variable typeset in a normal weight font represents a scalar; a bold weight font denotes a first- or second-order tensor. An overline indicates that the variable is defined with respect to the reference configuration and, if absent, the variable is defined with respect to the deformed configuration.

\subsection{Elastic forces}

\paragraph{Kinematics.}

\begin{figure}[t!]
  \centering
  \includegraphics[width=1.0\textwidth]{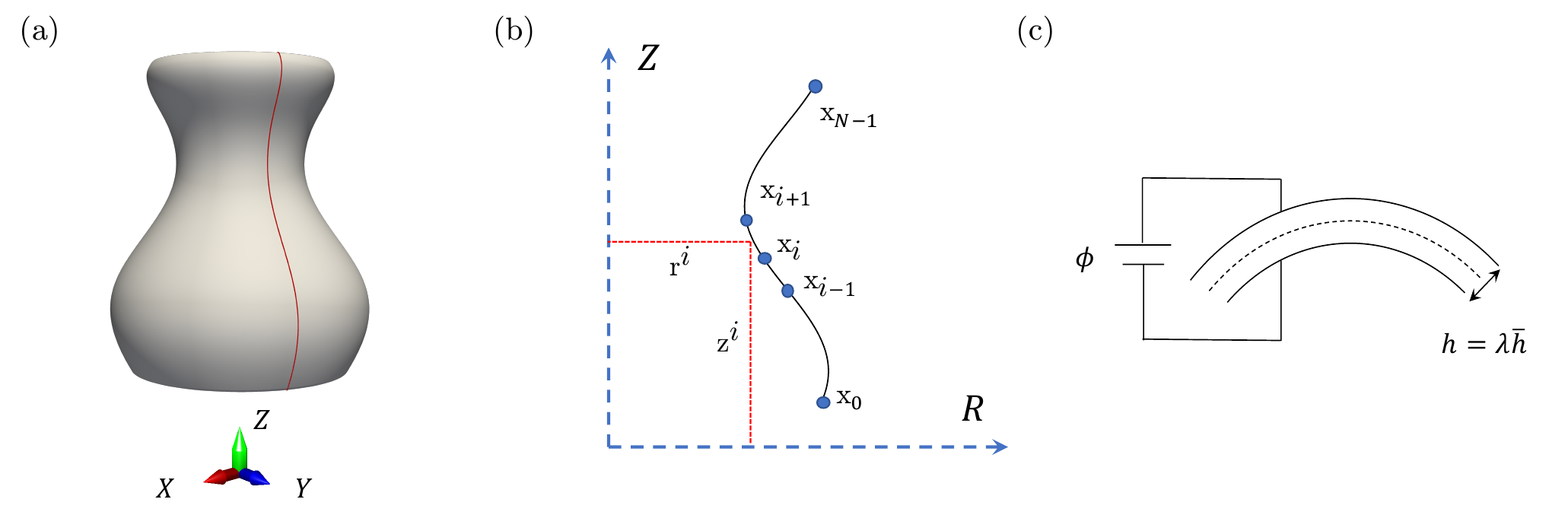}
  \caption{The three dimensional axisymmetric electroelastic membrane (a) can be generated by employing an order-reduced curve (b) and rotating it with respect to the $Z$ axis. (c) The membrane possesses an undeformed thickness $\bar h$ and a deformed thickness $h = \lambda \bar h$ when subjected to loads, including mechanical forces and electric potential difference between its top and bottom surfaces.}
  \label{fig:setupPlot}
\end{figure}

As shown in Fig.~\ref{fig:setupPlot}(a)) and Fig.~\ref{fig:setupPlot}(b), an axisymmetric membrane in $\mathcal{R}^3$ space can be represented by an order reduced curve in $\mathcal{R}^2$ space.
The axial direction is denoted as the $1$-direction, $R$, and the symmetric axis direction as the $2$-direction, $Z$.
Next, to describe its configuration, the rotational curvature is discretised into $N$ nodes, giving rise to a $2N$ sized degree of freedom vector $\mathbf{q}$ in $R-Z$ space, as  
\begin{equation}
\mathbf{q} = \{ \mathbf{x}_{0}, \mathbf{x}_{1}, ..., \mathbf{x}_{i}, ..., \mathbf{x}_{N-1} \},
\end{equation}
where $\mathbf{x}_{i} \equiv [ r_{i}, z_{i} ]$.
The edge vectors connecting the nodes are defined by,
\begin{equation}
\mathbf{e}^{i} = \mathbf{x}_{i+1} - \mathbf{x}_{i}.
\end{equation}
Its tangent and normal vectors satisfy,
\begin{equation}
\mathbf{t}^{i} =  {\mathbf{e}^{i} } / {l^{i}}
\quad \text{and} \quad
\mathbf{n}^{i} \cdot \mathbf{t}^{i} = 0.
\end{equation}
where $l^{i} = || \mathbf{e}^{i} ||$ is its length and is computed by the $\mathcal{L}_{2}$ norm of the edge vector.
Hereafter, subscripts are used to denote quantities associated with the nodes and superscripts for edges.
The average is used to transfer the node-based quantity to the edge-based quantity, e.g., the radius of $i$-th edge is denoted as 
\begin{equation}
r^{i} = \frac{1}{2} [r_{i}+r_{i+1}], 
\end{equation}
and the Voronoi length associated with the $i$-th node is given by,
\begin{equation}
l_{i} = \frac{1} {2} [{{l}}^{i-1} + {{l}}^{i} ].
\end{equation}
Next, the right Cauchy–Green deformation tensor for the axisymmetric membranes associated with the $i$-th edge is
\begin{equation}
\boldsymbol{C}^{i} =\operatorname{diag}\left([\lambda^{i}_1]^2, [\lambda^{i}_2]^2, [\lambda^{i}_3]^2\right),
\end{equation}
where the principal stretches $\lambda^{i}_1,~ \lambda^{i}_2,~\lambda^{i}_3$ are computed based on the geometry.
The first principal stretch of $i$-th edge along the meridional direction is,
\begin{equation}
\lambda^{i}_{1} =  { l^{i }} / { \bar{l}^{i}} ,
\end{equation}
and the second principal stretch of $i$-th edge along the circumferential direction is related to the expansion of the circle, i.e.,
\begin{equation}
\lambda^{i}_{2} =  { r^{i} } / { \bar{r}^{i} }.
\end{equation}
The third principal stretch of $i$-th edge is the change of the thickness,
\begin{equation}
\lambda^{i}_{3} = h^{i} / \bar{h}^{i},
\end{equation}
where $h$ is the thickness of the current edge element.
Also, we assume the thin membrane is volume conservative, such that 
\begin{equation}
\lambda^{i}_{1} \lambda^{i}_{2} \lambda^{i}_{3} = 1.
\end{equation}

\paragraph{Constitutive law.}

As the elastic membrane would experience a large expansion during the actuation process and the associated strain can exceed $100\%$, the hyperelastic constitutive law is required to describe its nonlinear behaviors. The elastic energy density for the $i$-th membrane element can be denoted as a function of three principal stretches formulated previously, i.e., $w^{i}(\lambda_{1}^{i},\lambda_{2}^{i},\lambda_{3}^{i})$. For example, the strain energy density per unit undeformed volume for a classical Mooney–Rivlin material model is expressed as
\begin{equation}
w^{i}(\lambda_{1}^{i},\lambda_{2}^{i},\lambda_{3}^{i}) = c_{1}[I_{1}^{i} - 3] + c_{2} [I_{2}^{i} - 3],
\label{eq:constitutiveEquation123}
\end{equation}
where two principal invariants $I_{1}^{i}$ and $I_{2}^{i}$ are computed as
\begin{equation}
\begin{aligned}
I_{1}^{i} &=  [\lambda^{i}_{1}]^2 + [\lambda^{i}_{2}]^2 + [\lambda^{i}_{1}]^{-2} [\lambda^{i}_{2}]^{-2} \\
I_{2}^{i} &= [\lambda^{i}_{1}]^{-2} + [\lambda^{i}_{2}]^{-2} + [\lambda^{i}_{1}]^{2}[\lambda^{i}_{2}]^{2}.
\end{aligned}
\end{equation}
The model can reduce to Neo-Hookean solid if $c_{2} = 0$.
The elastic energy of a 1D axisymmetric shell in discrete form is the sum of all stretching elements,
\begin{equation}
U^{\mathrm{ela}} = \sum_{i=0}^{N-2} \; w^{i} d\bar{v}^{i}.
\end{equation}
where $ d\bar{v}^{i} = 2 \pi \bar{r}^{i} \bar{l}^{i} \bar{h}^{i}$ is the volume of the $i$-th edge element.
Other types of nonlinear material models can also be adopted by simply replacing Eq.(\ref{eq:constitutiveEquation123}). 
The internal force vector and the associated local stiffness matrix are related to the variation (and the second variation) of the total potential,
\begin{equation}
\mathbf{F}^{\mathrm{ela}} = - \nabla U^{\mathrm{ela}} 
\quad \text{and} \quad
\mathbb{H}^{\mathrm{ela}} = \nabla \nabla U^{\mathrm{ela}},
\end{equation}
where $
\nabla(\bullet) =  {\partial (\bullet)} / {\partial \mathbf{q}}$ is the gradient operator and $\nabla \nabla(\bullet) $ refers to $  {\partial^2 (\bullet)} / {\partial \mathbf{q}^2} $.
The chain rule can be used to derive the closed-form expression of the elastic force vector and the stiffness matrix in terms of the degree of freedom vector $\mathbf{q}$.

\subsection{External forces}

\paragraph{Pneumatic actuation.}

Thin membranes under internal pressure are widely used in engineering applications and can be harnessed for actuation.
The equivalent force vector on the $i$-th node in the 1D membrane model, $\mathbf{x}_{i}$, subjected to a pressure difference between the inner and outer surfaces, is formulated as
\begin{equation}
\mathbf{f}_{p} = 2 \pi p \; r_{i} \; l_{i} \; \mathbf{n}_{i}.
\end{equation} 
The pressure force vector, $\mathbf{F}^{\mathrm{pre}} \in \mathbb{R}^{2N\times 1}$, can next be filled up based on the mapping, i.e., $\mathbf{f}_{p} \equiv [\mathbf{F}^{\mathrm{pre}}_{2i}, \mathbf{F}^{\mathrm{pre}}_{2i+1}]$.
The Jacobian for external pressure is given by taking the derivative of the force vector with respect to $\mathbf{q}$,
\begin{equation}
\mathbb{H}^{\mathrm{pre}} = -\nabla \mathbf{F}^{\mathrm{pre}}.
\end{equation} 
Note that the evolving pressure force needs to be updated based on the membrane current configuration instead of the reference shape.

\paragraph{Electrical actuation.}

An elastic membrane composed of a dielectric elastomer can also be actuated by the electric field, referring to Fig.~\ref{fig:setupPlot}(c).
In the current study, we assume that the electrical voltage is always along the membrane thickness direction, such that the free energy of $i$-th edge element is 
\begin{equation}
\psi^{i} = \frac {1} {2} \varepsilon 
 \left[\frac {\phi^{i}} {{h}^{i}} \right]^2 d \bar{v}^{i},
\end{equation}
where $ \varepsilon $ is the permittivity of the material, and $\phi^{i}$ is the electrical voltage between the $i$-th inner side and outer side of a membrane.
The internal force vector and the associated local stiffness matrix are related to the variation (and the second variation) of the total potential,
\begin{equation}
\mathbf{F}^{\mathrm{ele}} = - \sum_{i=0}^{N-2} \nabla \psi^{i}
\quad \text{and} \quad
\mathbb{H}^{\mathrm{ele}} = \sum_{i=0}^{N-2} \nabla \nabla \psi^{i},
\end{equation}

\paragraph{Contact force.}

Mechanical contact or indentation on shell structures is commonly applied in engineering applications \cite{liu2021effect}.
When a node, $\mathbf{x}_{i} $, forms contact with the other surface, a force is generated at the point.
The incremental potential contact method is employed to capture the nonlinear contact boundary between the deformable shell and a rigid surface ~\cite {li2020incremental}.
A smooth log-barrier potential with $C^2$ continuity is employed when the $i$-th node is within a critical distance with the rigid surface~\cite{li2020incremental},
\begin{equation}
\zeta_{i} =
\begin{cases}
- K_{c} \; [ d_{i} - \hat{d}]^2 \; \log( { d_{i} } / {\hat{d}}) \; &\mathrm{when} \; 0 \leq d_{i} < \hat{d}, \\
{0} \; & \mathrm{when} \; d_{i} \geq \hat{d},
\end{cases}
\label{eq:barrierContact0}
\end{equation}
where $d_{i}$ is the minimum distance between $i$-th node and the target object, $\hat{d}$ is the barrier parameter, and $K_{c}$ is the contact stiffness.
Note that the log-barrier force is zero when the distance is larger than $\hat{d}$. The repulsive interaction gradually increases as the distance decreases within $\hat{d}$. The repulsive force goes to infinite if $ d_{i} $ approaches zero.
The total contact potential is the sum of all contact elements, $\mathcal{C}$.
The gradient of the contact force vector can be derived analytically due to the $C^2$ continuity of the contact potential, i.e., the force and Jacobian can be formulated like the elastic potentials,
\begin{equation}
\mathbf{F}^{\mathrm{con}} = - \sum_{i \in \mathcal{C}} \nabla \zeta_{i} 
\quad \text{and} \quad
\mathbb{H}^{\mathrm{con}} = \sum_{i \in \mathcal{C}} \nabla \nabla \zeta_{i}.
\end{equation}

\subsection{Equations of motion}

Finally, the discrete equations of motion of a 1D membrane is formulated.
To obtain the equations of motion for the axisymmetric membranes, one constructs a time-invariant mass matrix, where the mass of the $i$-th node is expressed as
\begin{equation}
m_{i} = 2 \pi \bar{r}_{i} \; \bar{l}_{i} \; \bar{h}_{i} \; \rho,
\end{equation}
where $\rho$ is the material density.
The diagonal mass matrix $\mathbb{M} \in \mathbb{R}^{2N \times 2N}$ can be later implemented.
Moving forward, by using the force balance and the implicit Euler method, one can solve the discrete dynamic equations of motion and update the DOF vector from $t= t_{k}$ to $t= t_{k+1}$,
\begin{equation}
\mathcal{E} \equiv \mathbb{M} 
 \ddot{\mathbf{q}}(t_{k+1}) + \xi \mathbb{M} 
 \dot{\mathbf{q}}(t_{k+1}) - \mathbf{F}^{\text{ela}}(t_{k+1}) - \mathbf{F}^{\text{ext}}(t_{k+1}) = \mathbf{0}
\label{eq:EOM}
\end{equation}
with
\begin{equation}
\begin{aligned}
{\mathbf{q}}(t_{k+1}) &= {\mathbf{q}}(t_{k}) + \dot{\mathbf{q}}(t_{k+1}) \; \delta t \\
\dot{\mathbf{q}}(t_{k+1}) &= \dot{\mathbf{q}}(t_{k}) + \ddot{\mathbf{q}}(t_{k+1}) \; \delta t,
\end{aligned}
\end{equation}
where $ \delta t $ is the time step size, ${\xi}$ is the damping coefficient, and $\mathbf{F}^{\text{ext}}$ is the external force vector, e.g., pneumatic actuation, electrical actuation, and contact force, as discussed before.
At time step $t_{k+1}$, a new solution is first guessed on the basis of the previous state, i.e.,
\begin{equation}
\mathbf{q}^{(1)}(t_{k+1}) = \mathbf{q}(t_{k}) + \dot{\mathbf{q}}(t_{k}) \; \delta t.
\end{equation}
It is then optimised by utilizing the gradient descent algorithm, such that the new solution at the $(n+1)$-th step is,
\begin{equation}
\mathbf{q}^{(n+1)}(t_{k+1}) = \mathbf{q}^{(n)}(t_{k+1}) - [ {\mathbb{M}} / {\delta t^2} + \xi {\mathbb{M}} / {\delta t} + \mathbb{H}^{\text{ela}} + \mathbb{H}^{\text{oth}}] \backslash \mathcal{E}^{(n)}.
\label{eq:newtonMethod}
\end{equation}
Then, we update the current time step and move forward until the solution is within a prescribed tolerance.
Similar to the 1D rod model~\cite{huang2023discrete}, the Jacobian matrix for the 1D axisymmetric shell framework is a banded matrix, which can be solved in linear complexity.
This numerical framework is established for dynamic simulation. However, it can also be used for static equilibrium analysis under arbitrary loading and boundary conditions when damping is introduced into the system as an external dissipative force, and can automatically capture stable equilibrium configurations and avoid unstable equilibrium patterns if perturbations are added into the system \cite{han2003study,huang2023bifurcations}.
This framework is well-suited for the analysis of nonlinear behaviors, e.g., buckling and snapping \cite{huang2023exploiting}.

\section{Model validation}\label{sec:Modelvalidation}
\subsection{Hyperelastic membranes}
\label{sec:hyperModelvalidation}

In this section, the analysis of hyperelastic membranes modeled by 1D axisymmetric DDG elements is conducted. It includes three types of hyperelastic membranes: a circular plate, a spherical balloon, and a toroidal membrane. The Mooney-Rivlin constitutive relation is employed to characterize the hyperelastic behavior. The results obtained will be compared with alternative numerical methods as well as analytical solutions in order to validate the proposed method.

\subsubsection{Pressurising a circular plate}
\label{sec:cicrcular_plate}

\begin{figure}[t!]
  \centering
  \includegraphics[width=1.0\textwidth]{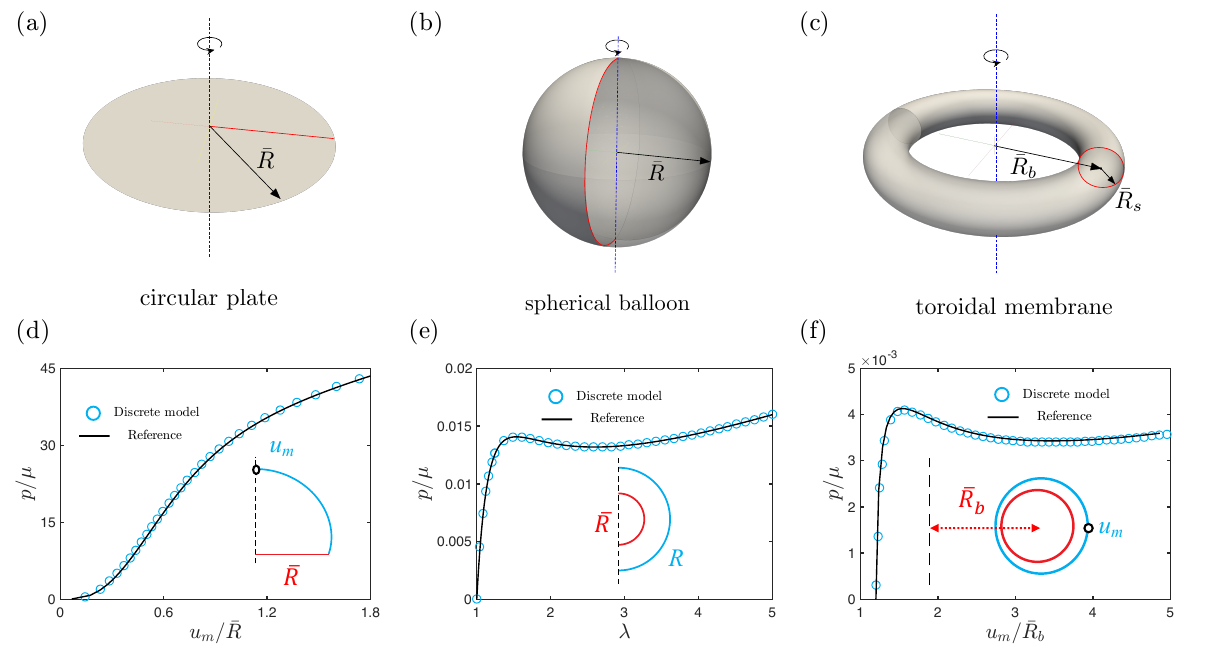}
  \caption{Validations of three example hyperelastic structures: (a) The inflation of a circular plate with radius $\bar R = 7.5$ m and reference results are presented in~\cite{liu2023computational}. (b) A spherical balloon with radius $\bar R=10$ m and thickness $\bar h=0.1$ m and analytical solution is shown in~\cite{holzapfel2002nonlinear}. (c) The inflation of a toroidal membrane with outer radius $\bar{R}_{b} = 10$ m, inner radius $\bar{R}_{s} = 2$ m, and thickness $\bar h= 0.1$ m, with reference solutions generated using the approach outlined in~\cite{liu2020coupled}. The normalised pressure $\hat p$ is plotted against (d) 
  the maximum normalised displacement during the pressurisation of the circular plate, (e) the stretch during the inflation of the spherical balloon, and (f) maximum normalised displacement during the inflation of the toroidal membrane.}
  \label{fig:hyperDemoPlot}
\end{figure}

Initially, a pressurized hyperelastic plate, simply supported on its circular boundary, is examined.
The pressure is considered a uniform out-of-plane load, consistently perpendicular to the deformed plate.
This classic example problem has been extensively studied in the literature~\cite{oden1970analysis,hughes1983nonlinear,Cirak:2001aa,nama2020nonlinear,liu2023computational}. Fig.~\ref{fig:hyperDemoPlot}(a) illustrates that the circular plate is modeled by revolving a line segment. The axis of rotation is perpendicular to the line, and one end of the line is situated on this axis. A circular plate with an undeformed radius $\bar{R} = 7.5$ m and an undeformed thickness $\bar{h} = 0.5$ m is under consideration, and its material properties are characterized by the Mooney–Rivlin model. Here, we define dimensionless variables as
\begin{equation}
    \hat{p} = \frac{p}{\mu}, \; \hat{c}_1 = \frac{c_1}{\mu}, \; \hat{c}_2 = \frac{c_2}{\mu},
\end{equation}
where $\hat{p}$ denotes the normalised pressure, while $\hat{c}_1$ and $\hat{c}_2$ are two material parameters, which are set as $0.4$ and $0.1$, respectively, in this example. The numerical results from the proposed method are shown in Fig.\ref{fig:hyperDemoPlot}(d), illustrating the relationship between the normalised maximum displacement $u_m/\bar{R}$ and $\hat{p}$, which signifies the pressure applied perpendicular to the plate. A comparative analysis is conducted against the results presented in both~\cite{nama2020nonlinear} and~\cite{liu2023computational}, revealing a perfect agreement among the results.

\subsubsection{Spherical balloon}
\label{sec:hyperModelvalidation_balloon}

The second case considered is the inflation of a spherical balloon, which, was previously examined analytically in~\cite{holzapfel2002nonlinear}. This example serves as a widely used benchmark for evaluating incompressible hyperelastic membrane and shell formulations~\cite{Cirak:2001aa,chen2014explicit,kiendl2015isogeometric}. The inflation process involves the application of uniform pressure from the inside. The construction of the undeformed spherical balloon involves the rotation of a half-circle, as shown in Fig.~\ref{fig:hyperDemoPlot}(b). The symmetrical axis is defined as the line connecting the two endpoints of the half-circle. For an incompressible Mooney--Rivlin constitutive model, the analytical expression for the normalized internal pressure $p$ of the uniformly inflated balloon is given by
\begin{equation}
\hat{p} = \frac{ 4 \bar h}{\bar R} \left[\hat{c}_1 [\lambda^{-1} - \lambda^{-7}] - \hat{c}_2[\lambda^{-5} - \lambda]\right],
\end{equation}
$\bar R$ represents the undeformed radius of the spherical shell, and $\bar h$ is the undeformed thickness. It is noteworthy that if $\hat{c}_2$ is set to zero, the model simplifies to the neo-Hookean model. For the spherical balloon, the two in-plane stretches, $\lambda_1$ and $\lambda_2$, are equal during inflation. Thus, one parameter, $\lambda = R/\bar{R}$, is used to represent the in-plane stretch. A spherical balloon with an undeformed radius $\bar R = 10$ m and a thickness $\bar h = 0.1$ m is considered. The material parameters $\hat{c}_1 = 0.4375$ and $\hat{c}_2 = 0.0625$, thus $\hat{c}_1/\hat{c}_2 = 7$. 
A numerical assessment is carried out using the current method, and the obtained result is compared against the analytical solution. The numerical results exhibit precise alignment with the analytical solution~\cite{holzapfel2002nonlinear}
shown in Fig.~\ref{fig:hyperDemoPlot}(e).

\subsubsection{Inflation of a toroidal membrane}
\label{sec:hyperModelvalidation_torus}

The third example considered is the inflation of a toroidal membrane. A toroidal membrane can be characterised as a surface formed by the rotation of a circle around an axis within the same plane, with the condition that the axis does not intersect the circle, illustrated in Fig.\ref{fig:hyperDemoPlot}(c). A torus is the simplest example of a genus 1 orientable surface, which exhibits unique geometric and mechanical characteristics that make them suitable for various applications, such as tires, air springs, soft grippers, and actuators. We consider a  toroidal membrane with an undeformed radius of the circular cross-section  $\bar{R}_s = 2$ m. The outer radius, the distance from the center of the cross-section to the center of the torus, is $\bar{R}_b = 10$ m, while the thickness is $\bar{h} = 0.1$ m. The material parameters are $\hat{c}_1 = 0.4375$ and $\hat{c}_2 = 0.0625$. The inflation of the toroidal membrane is simulated using the current method, and the obtained results are compared with the alternative FE-based approach detailed in~\cite{liu2023computational}. Fig.~\ref{fig:hyperDemoPlot}(f) illustrates the variation of normalised pressure $\hat{p}$ with normalised maximum displacement $u_m/\bar{R}_b$ for both methods, showing a high level of agreement.

\subsection{Electroelastic membranes}
\label{sec:hyperDeModelvalidation}

This section aims to further validate the current method for analysing axisymmetric electroelastic membranes. We consider an electroelastic membrane with hyperelastic property but also responding to electrical actuation. The application of an electric potential difference across the thickness is anticipated to result in  uniform stretching of the membrane, thereby inducing mechanical softening. Thus, the electrical loads will not just be viewed as external loads but will also alter the stiffness of the membrane in response to mechanical force. The verification is achieved by examining the response of both mechanical force and electrical actuation applied to the  two latter cases considered in Section~\ref{sec:hyperModelvalidation}: spherical balloon and toroidal membrane.

\begin{figure}[b!]
  \centering
  \includegraphics[width=0.85\textwidth]{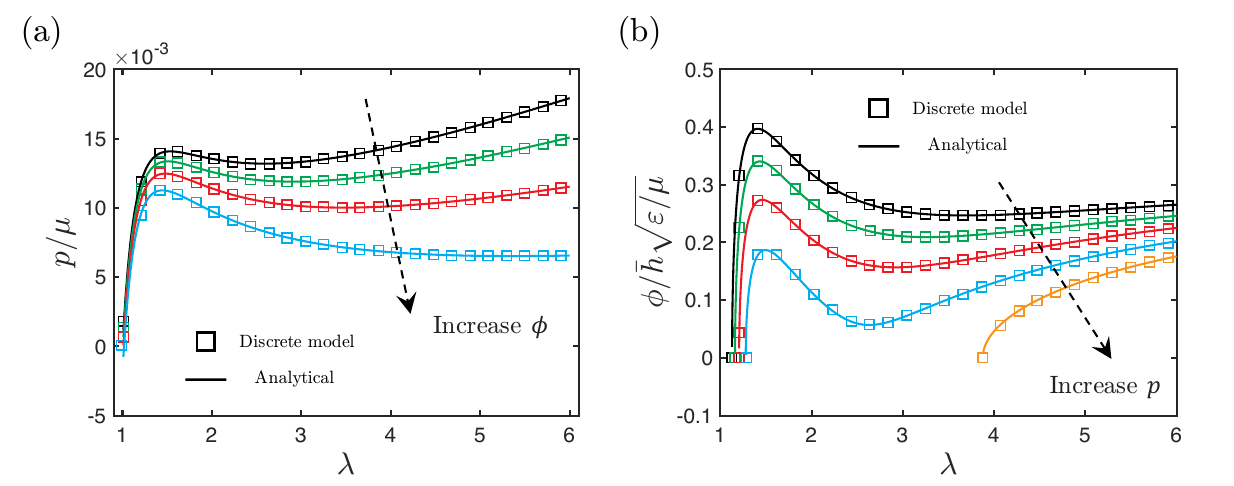}
  \caption{An electroelastic spherical balloon with radius $\bar{R} = 10$ m, thickness $\bar{h} = 0.1$ m, and $\hat{c}_1/\hat{c}_2 = 7$. (a) Variation of normalized pressure $\hat{p}$ over the stretch $\lambda$. The electrical load $\hat{E}$ varies within a set of values ${0}$, ${0.1538}$, ${0.2308}$, and ${0.3077}$. (b) Variation of electrical load over the stretch $\lambda$. The normalized pressure $\hat{p}$ varies within a set of values $\{9.5, 10.7, 11.8, 13.0, 14.2\} \times 10^{-3}$. Here, the symbols represent the discrete simulation, and the lines are from the analytical solution. 
  }
  \label{fig:hyperDeballDemoPlot}
\end{figure}

\subsubsection{Electroelastic spherical balloon}

The mechanical response of the spherical balloon under electrical actuation is examined by imposing an electric potential difference across its thickness. The incompressible Mooney-Rivlin constitutive relation is employed to characterise its hyperelastic behaviour. The identical problem configuration, including both geometric and material parameters as detailed in Section~\ref{sec:hyperModelvalidation_balloon}, is utilised. The analytical solution for the pressure inside the electroelastic spherical balloon~\cite{holzapfel2002nonlinear} is derived as
\begin{equation}
\hat{p} = \frac{ 4 \bar h}{\bar R} \left[\hat{c}_1 [\lambda^{-1} - \lambda^{-7}] - \hat{c}_2[\lambda^{-5} - \lambda]- \frac{1}{2}\hat{E}^2\lambda\right].
\end{equation}
where $\hat{E}$ denotes a dimensionless electrical load, which can be explicitly defined as
\begin{equation}
 \hat{E}  = \frac {\phi} {\bar{h}} \sqrt{ \frac {\varepsilon} {\mu} } 
\end{equation}
Fig.~\ref{fig:hyperDeballDemoPlot}(a) illustrates the in-plane stretch parameter $\lambda$ plotted against the normalized pressure $\hat{p}$ for the chosen electrical loads, i.e., $\hat{E} = $ $0$, $0.1538$, $0.2308$, and $0.3077$. The numerical results agree well with the analytical solutions. The results indicate that the applied electrical load reduces the peak pressure of the spherical balloon, making the membrane less resistant to mechanical loads. For small deformations, the membrane demonstrates significant stiffness when subjected to inflation. The ratio of the pressure to stretch is large initially, and this ratio remains consistent across various electrical loads. With the increase in electrical load, the limit point occurs at a similar volume change, but the associated pressure at the peak notably diminishes. After the peak, the pressures all decrease with inflation but subsequently increase again with large deformations, illustrating a distinct snap-through path. However, the applied electrical load elongates the snap-through path.

Fig.~\ref{fig:hyperDeballDemoPlot}(b) similarly plots the stretch against the normalized electrical load for balloons of varying internal pressures. The sequentially chosen values as the controlled internal pressures $\hat{p}$ are ${\{9.5, 10.7, 11.8, 13.0, 14.2\}} \times 10^{-3}$. The first four pressures reside below the limit point of the hyperelastic spherical balloon ($\hat{E} = 0$), while the final case surpasses the limit point. The spherical balloon is first inflated to the specific pressure, after which an electrical load is applied across its thickness. As deformation progresses, the applied electrical load approaches a limit point. After the limit point, the electrical load diminishes as the deformation increases and will rise again after large deformation. For $\hat{p} = 13.0 \times 10^{-3}$, a relatively short snap-through path is observed, indicating that the applied electrical load can induce snap-through instability. Quantitative agreements between the predictions from the discrete model (line) and the results from the analytical formulation (symbol) indicate the accuracy of our model.

Upon examination of both figures, it is evident across all curves that the normalized pressure or electrical load decreases after reaching their limit points.
However, certain curves subsequently exhibit an increase shortly thereafter.
In such instances, as the load approaches the limit point, the balloon will undergo a sudden expansion, transitioning to a new stable state without increasing pressure or electric load.
This snap-through instability phenomenon in hyperelastic balloons, characterized by a sudden volume change triggered by internal pressure, is widely recognized in literature~\cite{crisfield1981fast,pecknold1985snap}.
However, for dielectric balloons, the instability can also be induced by applying electrical load~\cite{li2013giant}.
Therefore, this phenomenon can be utlised in the design of electrical actuators for applications in soft robotics.

\begin{figure}[b!]
  \centering
  \includegraphics[width=0.85\textwidth]{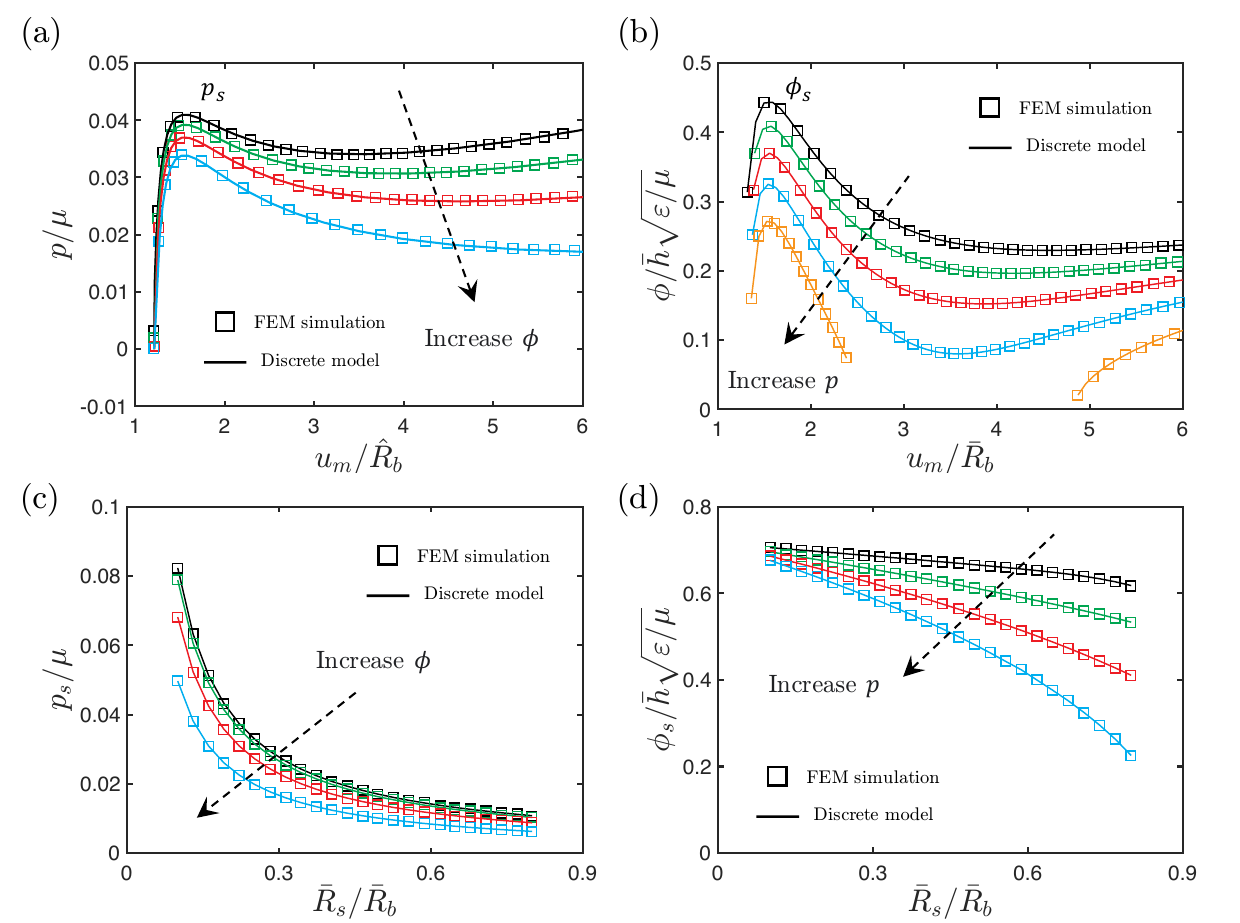}
  \caption{An electroelastic toroidal membrane with outer radius $\bar{R}_b = 10$ m, inner radius $\bar{R}_s = 2$ m, thickness $\bar{h}=0.1$ m, and $\bar c_1/\bar c_2 = 7$, is examined. (a) Variation of normalised pressure $\hat p$ over the maximum normalised displacement $u_m/\bar R_b$. The electrical load $\hat{E}$ varies within a set of values $\{ 0, 0.1538, 0.2308, 0.3077 \}$. (b) Variation of electrical load over the maximum normalised displacement $u_m/\bar R_b$. The normalised pressure $\hat{p}$ varies within $\{ 0.0260, 0.0284, 0.0308, 0.0331, 0.0355 \}$. (c) Variation of the normalised pressure at limit point $\hat p_s$ over the aspect ratio $\bar R_s/\bar R_b$, while the electrical load $\hat{E}$ varies within $\{0, 0.1538, 0.3077, 0.4615 \}$. (d) Variation of electrical load at limit point over the aspect ratio $\bar R_s/\bar R_b$, while the normalised pressure $\hat{p}$ varies within $\{0.0260, 0.0284, 0.0710, 0.0947 \}$. Here, the symbols are from 3D FEM analysis and the lines are from the discrete simulation.}
  \label{fig:hyperDetoriDemoPlot}
\end{figure}

\subsubsection{Inflation of electroelastic toroidal membrane}

The electroelastic coupling behavior of the toroidal membrane is also investigated, and the 3D FEM simulation is used for cross-validation \cite{liu2023computational}. Here, the problem setup, including both geometric and material parameters, is identical to the example in Section~\ref{sec:hyperModelvalidation_torus}, and an electric potential difference is similarly applied across the thickness to alter the mechanical behavior of the toroidal membrane. Fig.\ref{fig:hyperDetoriDemoPlot}(a) illustrates the maximum normalized displacement $u_m/\bar R_b$ against the normalized pressure $\hat{p}$ for the applied electrical loads $\hat{E} \in {0, 0.1538, 0.2308, 0.3077}$. Comparative analysis is conducted with an alternative approach detailed in\cite{liu2020coupled}, where a thorough investigation of such membranes has been undertaken. The two numerical results exhibit close alignment, thereby validating the proposed method. Similar to the electroelastic spherical balloon, the plots indicate that the applied electrical load diminishes the limit point pressure $p_s$ and weakens the ability of the toroidal membrane to resist mechanical loads. Fig.~\ref{fig:hyperDetoriDemoPlot}(b) plots the maximum normalized displacement $u_m/\bar R_b$ against the electrical load $\hat{E}$ for controlled internal pressure $\hat{p}$ varying within a set of values ${ 0.0260, 0.0284, 0.0308, 0.0331, 0.0355 }$. In these instances, as the toroidal membranes inflate, the electrical loads approach limit points and decrease notably afterward. Therefore, the impact of geometry on the limit point pressure is further investigated.

An essential parameter for a torus is its aspect ratio, defined as the ratio of the major radius (distance from the center of the cross-section to the symmetric axis) to the minor radius (radius of the circular cross-section), denoted as $\bar{R}_s/\bar{R}_b$. Fig.\ref{fig:hyperDetoriDemoPlot}(c) illustrates the relationship between the limit point pressure and the aspect ratio for electrical loads $\hat{E}$ varying within a set of values ${0, 0.1538, 0.3077, 0.4615}$. As the aspect ratio increases, the limit point pressure diminishes and approaches a convergence value. Moreover, in terms of response to internal pressure, the influence of the electrical load on a toroidal membrane with a smaller aspect ratio is more pronounced than on one with a larger aspect ratio. Conversely, when considering the response to the electrical load, the influence of internal pressure on a toroidal membrane with a larger aspect ratio is more pronounced than on one with a smaller aspect ratio. Fig.\ref{fig:hyperDetoriDemoPlot}(d) showcases the association between the limit point electrical load and the aspect ratio, considering internal pressure values $\hat{p}$ varying within a set of values ${0.0260, 0.0284, 0.0710, 0.0947}$. Quantitative agreements between the predictions from the discrete model (line) and the results from the 3D FEM analysis (symbol) indicate the effectiveness of our model.

\section{Robotic applications}\label{sec:Roboticapplication}

Based on the geometric characteristics and electroelastic coupling behaviors of circular plates, spherical balloons, and toroidal membranes, we propose three robotic applications: a soft circular fluid pump, a soft gripper utilizing the toroidal membrane, and a soft actuator designed in the shape of a spherical balloon.

\subsection{A soft circular fluid pump}

\begin{figure}[b!]
  \centering
  \includegraphics[width=1.0\textwidth]{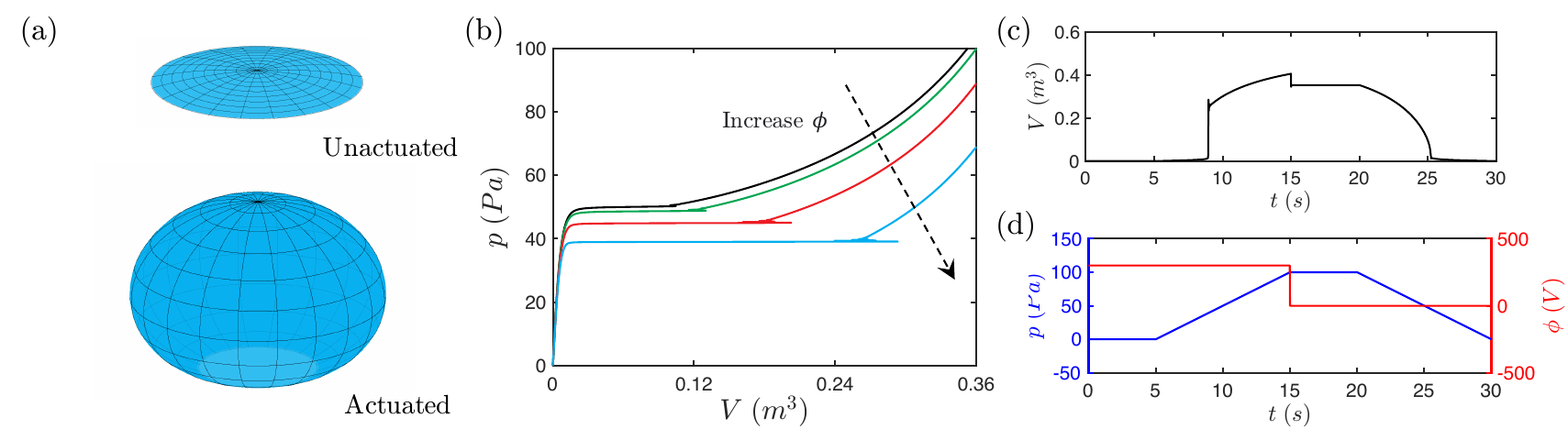}
  \caption{A soft fluid pump utilizing an electroelastic circular membrane, with radius $\bar{R} = 0.1$ m and thickness $\bar{h}=10^{-4}$ m. (a) Illustrations are provided of the unactuated and actuated states of the fluid pump. (b) The relationship between the enclosed volume and the fluid pressure inside the membrane is shown. (c) Temporal evolution of the enclosed volume. (d) The temporal evolution of the inside fluid pressure and the electric potential applied to the membrane.}
  \label{fig:planeActPlot} 
\end{figure}

The snap-through phenomenon offers a promising mechanism for achieving substantial volume change within electroelastic membranes, making this an attractive approach for efficient large-volume fluid pumping. Recently, a novel operating mechanism has been developed by Li et al.~\cite{li2017mechanism} to achieve a reversible snap-through effect for efficient large-volume fluid pumping using a circular diaphragm made of dielectric elastomer. They later introduced an improved design of a dual-membrane dielectric elastomer pump, incorporating an active dielectric elastomer membrane that interacts with a passive elastic membrane to ensure that the pump can snap under small pressures~\cite{li2017robust}. In this work, we employ the proposed 1D DDG-based method to conduct a numerical investigation of this pumping mechanism.

The identical  configuration described in Section~\ref{sec:cicrcular_plate} is employed, and the unactuated and actuated circular membranes are illustrated in Fig.\ref{fig:planeActPlot}(a). However, the circular plate with the Mooney--Rivlin constitutive model does not exhibit the snap-through phenomenon under the specified constraints and loading conditions. Thus, the Gent constitutive relationship~\cite{gent1996new,suo2010theory} is utilized, which can be described as
\begin{equation}
w^{i}(\lambda_{1}^{i},\lambda_{2}^{i},\lambda_{3}^{i}) = - \frac {\mu J_\mathrm{lim}} {2} \log \left[1 - \frac{[I^i_1 -3]}{J_\mathrm{lim}}  \right] ,
\label{eq:constitutiveEquation}
\end{equation}
where $\mu=45.0$ KPa, $J_\mathrm{lim}$ denotes the volume deformation limit of the material, which is set at $270$. The fluid at the bottom of the membrane gradually pressurizes the membrane at a rate of $\dot{p} = 10$ Pa/s, and the material density $\rho = 1000$ kg/m$^3$ and the damping coefficient $\xi = 1.0$ are used.

Fig.\ref{fig:planeActPlot}(b) plots the relationship between the volume of the enclosed fluid and the internal pressure. With the increase in electric potential $\phi$, ranging from $0$ to $300$ V in intervals of $100$ V (where the permittivity $\varepsilon$ is assumed to be $10^{-10}~\mathrm{C}\cdot (\mathrm{Vm})^{-1}$), the snap-through occurs earlier, accompanied by a longer path. This indicates a substantial increase in the volume change resulting from the application of electric potential, which showcases the effectiveness of dielectric elastomers in functioning as fluid pumps. Subsequently, a complete pumping mechanism, incorporating both fluid inflow and outflow, is simulated and illustrated in Fig.\ref{fig:planeActPlot}(c). An electric potential $\phi = 300$ V is applied to the circular membrane at $t = 0$ s, but it does not actuate the membrane. When $t = 5$ s, the membrane starts to experience pressurization due to fluid inflow at a rate of $\dot p = 10$ Pa/s, leading to snap-through at $t = 8.9$ s. The pressurization ceases at $t = 15$ s, concurrently with the withdrawal of the electric potential. At this moment, there is a sudden loss of volume, measured at about $0.3$ m$^3$. At $t = 20$ s, the membrane initiates depressurization caused by fluid outflow at $\dot p = -10$ Pa/s. Without any applied electric potential, the enclosed volume gradually decreases and undergoes a snap back at $t = 25$ s. The circular membrane returns to its unactuated state at $t = 30$ s.

\begin{figure}[t!]
  \centering
  \includegraphics[width=1.0\textwidth]{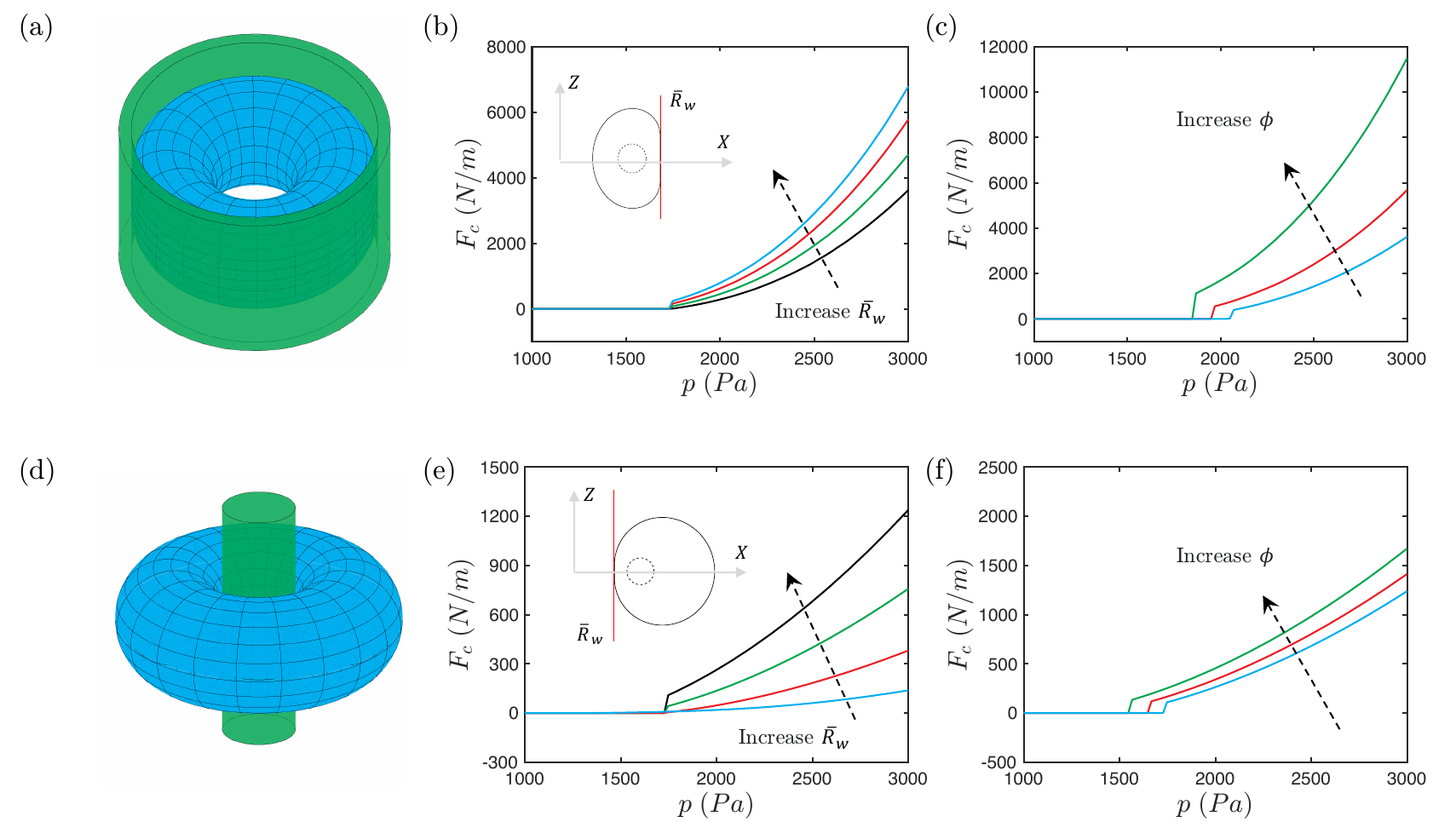}
  \caption{A soft gripper utilizing an electroelastic toroidal membrane, with outer radius $\bar{R}_b = 0.1$ m, inner radius $\bar{R}_s = 0.02$ m, thickness $\bar{h} = 10^{-4}$ m, shear modulus $\mu = 4.225 \times 10^5$ Pa, and $c_1/ c_2 = 7$. Case 1: (a) Illustration of a toroidal gripper using its inner region to securely grasp a cylinder. (b) The relationship between contact force and internal pressure is examined as the radius of the cylinder, denoted by $\bar R{w}$, varies within the set values of $\{0.04, 0.05, 0.06, 0.07\}$ m, in which no electric potential difference is applied. (c) The correlation between contact force and internal pressure is explored using a cylinder of constant radius $\bar R{_w} = 0.04$ m, however, $\phi$ varies within $\{0, 1000, 1500\}$ V. Case 2: (d) Illustration of a toroidal gripper for grasping a cylindrical shell. (e) The relationship between contact force and internal pressure is examined as the inner radius of the cylindrical shell, denoted by $\bar R_{w}$, varies within $\{0.13, 0.17, 0.21, 0.25\}$ m, in which no electric potential difference is applied. (f) The correlation between contact force and internal pressure is explored using a cylindrical shell with inner radius $\bar R_{w} = 0.25$ m, and $\phi$ varies within $\{0, 1000, 1500\}$ V.}
  \label{fig:toriDemoPlot}
\end{figure}

\subsection{A soft toroidal gripper}

A soft gripper utilizing an electroelastic toroidal membrane is proposed. The torus, characterized as a genus 1 orientable surface, inherently possesses a central hole within its geometric structure, making it ideal for a gripper. Fig.~\ref{fig:toriDemoPlot}(a) shows a toroidal gripper encircling a cylindrical rod within its interior region. The toroidal membrane has an outer radius of $\bar{R}_s = 0.1$ m and an inner radius of $\bar{R}_b = 0.02$ m with a thickness of $\bar{h} = 0.1$ mm. The shear modulus of the material is $\mu=4.225\times 10^5$ Pa, and $c_1$ and $c_2$ are $0.4375\mu$ and $0.0625\mu$, respectively. The cylindrical rod shares a symmetrical axis with the toroidal membrane. Owing to its significantly greater stiffness, it is considered a rigid body. Initially, there exists a gap between the rod and the toroidal membrane. As the pressure within the toroidal gripper increases, the enclosed membrane undergoes snap-through behavior characterized by significant deformation. Its interior region diminishes, resulting in contact with the rod. Upon the escalation of the internal pressure, the contact force also increases, causing the gripper to firmly grasp the rod.

We first consider the toroidal gripper as a pure elastic membrane with no electrical load applied.
In Fig.~\ref{fig:toriDemoPlot}(b), the plot illustrates the relationship between pressures inside the toroidal gripper and the contact forces for grasping rods of different sizes.
The radius of the cylindrical rod is denoted as $\bar{R}_w$, and the chosen set of values is ${0.04, 0.05, 0.06, 0.07 }$ m.
When grabbing a relatively slender rod with $\bar{R}_w = 0.04$ m, there initially exists a gap of $0.04$ m between the rod and the inner circumference of the toroidal membrane.
The gap allows the toroidal membrane to snap and inflate considerably, enabling the gripper to achieve sufficient contact force to securely grasp the rod.
As the internal pressure escalates to the limit point (approximately $1750$ Pa), there is a pronounced increase in the contact force.
When the identical toroidal gripper is adopted to grasp increasingly stouter rods, progressively reduced contact forces are observed.
Upon selecting a rod with a radius of $\bar{R}_w = 0.04$ m, one applies an electric potential difference across the thickness of the membrane; the chosen values for $\phi$ are $1000$ V and $1500$ V."
The application of electric potential causes the toroidal membrane to snap-through at a reduced pressure, leading to increased contact forces as the value of $\phi$ escalates, as shown in Fig.~\ref{fig:toriDemoPlot}(c).
Hence, the gripper can be activated by adjusting the electric potential difference to initiate contact and securely grasp the objects.
It is noteworthy that in all the aforementioned cases, the toroidal gripper requires extremely large volume expansion to effectively grasp the rods, which could potentially induce other instabilities.

Fig.~\ref{fig:toriDemoPlot}(d) showcases an alternative application of the toroidal membrane, wherein it functions as a gripper for cylindrical shells.
Furthermore, it possesses the capability to be adaptably used on arbitrary objects that possess internal circular voids.
In this particular application, the toroidal membrane is initially positioned within the cylindrical shell and subsequently inflates to establish contact with the inner surface.
Owing to the restraint imposed by the cylindrical shell, the toroidal membrane can not undergo large deformation.
The cylindrical shell is considered a rigid body.
In Fig ~\ref{fig:toriDemoPlot}(e), the plot illustrates the relationship between pressure inside the toroidal membrane and the corresponding contact forces applied on the inner surfaces of cylinders of different sizes.
The parameter $\bar R_w$ represents the inner radius of the cylindrical shell in this case.
Four cylindrical shells are chosen, with $\bar R_w$ values varying within the set ${0.13, 0.17, 0.21, 0.25}$ m, and the toroidal gripper as a pure elastic membrane is also considered without any electrical load applied.
For a cylindrical shell with a radius of $0.13$ m, the membrane inflates to its limit point, and, it begins to establish contact with the inner surfaces and the membrane does not undergo a snap-through transition.
Thus a smooth curve is observed.
As the dimensions of the cylindrical shell increase, the toroidal membrane has more space and can exhibit sudden volume changes when inflated to its limit pressure, resulting in increased contact force.
Fig.\ref{fig:toriDemoPlot}(f) demonstrates that, with a specified cylindrical shell of $\bar R_w = 0.25$ m, gradually adjusting the electric potential difference to $1000$ V and $1500$ V reduces the limit pressure of the toroidal membrane.
Consequently, the sudden volume of the toroidal membranes occurs at an earlier stage, leading to an increase in the contact force with the elevation of the electric potential difference.
Therefore, in this scenario, the gripper can also be activated by applying the electric potential difference to establish contact and firmly grasp the objects.

\begin{figure}[t!]
  \centering
  \includegraphics[width=1.0\textwidth]{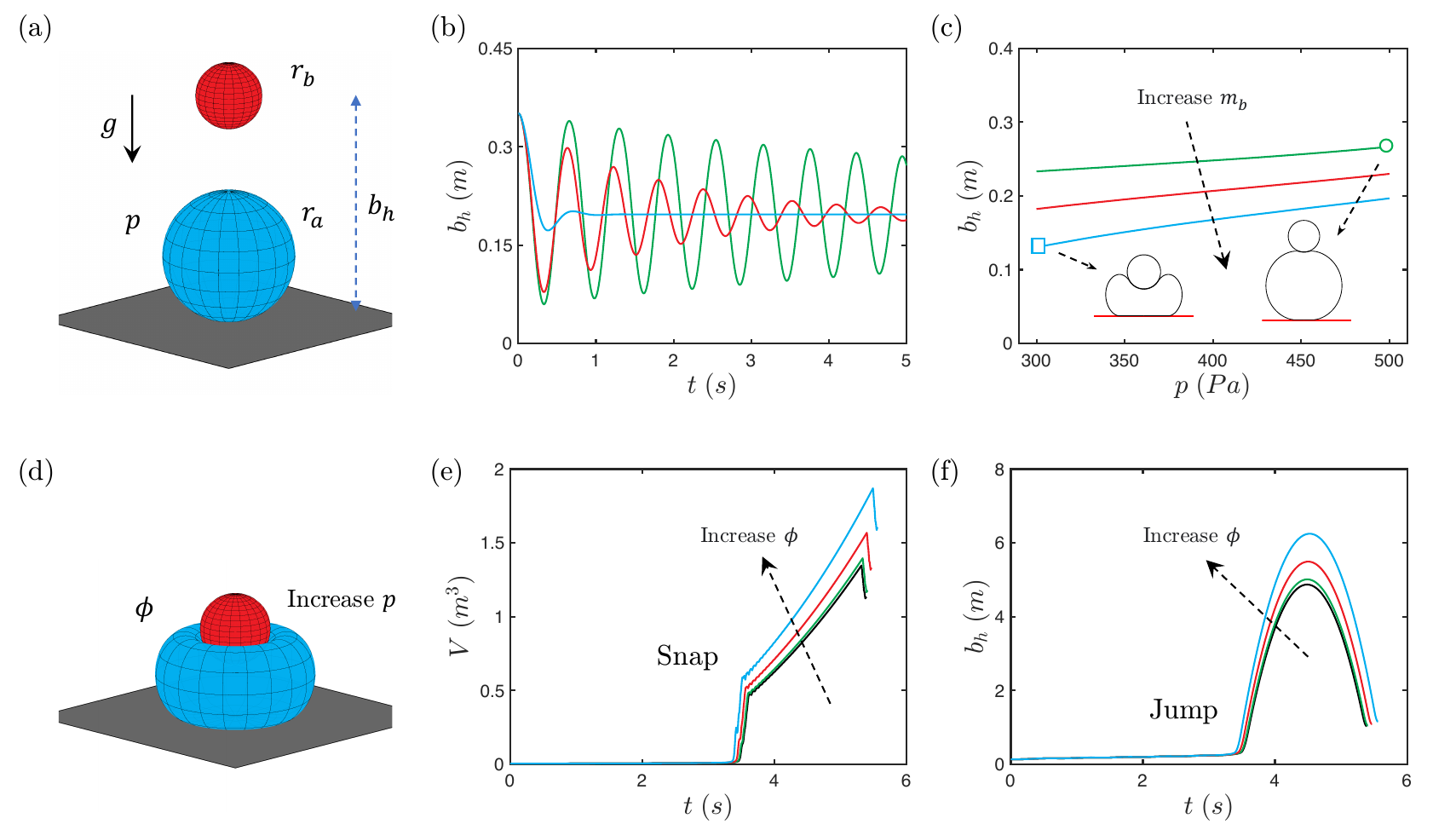}
1  \caption{A soft actuator utilizing an electroelastic spherical membrane for ball actuation. A spherical membrane, colored blue, with a radius denoted as $r_{a}$ upheld by a rigid plate. The thickness of the membrane is $\bar{h}=10^{-4}$ m and shear modulus denotes $\mu = 4.225 \times 10^5$ Pa with $c_1/c_2 = 7$. A rigid ball, colored red and possessing a radius of $r_{b}$ is employed to simulate dynamic and contact interactions with the membrane. Case 1: (a) The ball with $r_{a} = 0.1$ m is positioned above the spherical membrane at a distance from the plate denoted as $b_{h}$. Upon release, it will accelerate under the influence of gravity, where gravitational constant  $g= -10$ m/s$^2$. The spherical membrane functions as a buffer, mitigating movement.
(b) Variations in ball height $b_{h}$ are plotted over time for different damping ratios, varying in $\{0.1, 1.0, 10.0 \}$, with the spherical membrane maintained at an internal pressure of $p=500$ Pa. The ball with $r_{b} = 0.05$ m, weighing $m_{b} = 0.5$ kg.(c) Variations in ball height at equilibrium state are plotted over internal pressures for different masses, $m_{b}$ varies in a set of $ \{0.1, 0.3, 0.5 \}$ kg. Case 2: (d) The ball initially contacts with the spherical membrane at an equilibrium state with time $t = 0$ s. The spherical membrane undergoes a inflation with an increasing rate of $ \dot{p} = 100$ Pa/s. The ball will be induced to execute a vertical jump, subsequently descending back onto the membrane. (e) Time-dependent variations in enclosed volume are plotted for different electric potential differences, {$\phi$ varies in $\{0, 200, 400, 600\}$ V.} (f) Time-dependent variations in ball height are plotted corresponding to different electric potential differences, $\phi$ varies in $\{0, 200, 400, 600\}$ V. }
  \label{fig:ballDemoPlot}
\end{figure}

\subsection{A soft spherical actuator}

An inflated electroelastic spherical membrane is utilized in the design of a soft actuator. 
This spherical configuration exhibits central symmetry, ensuring the even distribution of stresses throughout the membrane.
Fig.~\ref{fig:ballDemoPlot}(a) shows the setup of the soft actuator.
The spherical membrane with radius $r_b$ is inflated initially with internal pressure $p$ and positioned onto a rigid plate.
A rigid ball with a mass $m_b$, aligned along the same vertical symmetric axis as the spherical membrane, is initially located above the membrane.
Upon releasing the ball, it will descend under the influence of gravity, where $g$ denotes the gravitational constant, and subsequently come into contact with the spherical membrane.
Firstly, the spherical membrane functions as a buffer, absorbing the momentum imparted by the descending ball.
Owing to the damping effect from the spherical membrane, the ball will undergo dynamic oscillations before eventually attaining a state of equilibrium.
Fig.~\ref{fig:ballDemoPlot}(b) demonstrates the variation in the height of the ball, measured from the center of the ball to the plate, with respect to time.
Three scenarios are examined, corresponding to damping ratio $\xi$ with values of $0.1$, $1.0$ and $10.0$, respectively.
All these scenarios will ultimately achieve a state of equilibrium with the same $b_h$.
However, the duration required to achieve this equilibrium varies among the scenarios.
It is noteworthy that the scenario with a higher damping ratio will attain equilibrium in a shorter duration.
Fig.~\ref{fig:ballDemoPlot}(c) showcases the variation in the height at equilibrium state $b_h$ with respect to increasing internal pressure, considering varying masses of the ball, specifically $0.1$, $0.3$ and $0.5$ kg.
As expected, the elevation of the internal pressure will counteract the deformation induced by the ball, resulting in a greater $ b_h$.
Conversely, an increase in the ball's mass corresponds to a reduction in $b_h$.

Subsequently, the inflation of the electroelastic spherical membrane to activate the ball is examined, considering various applied electrical loads.
A ball with a mass of $0.5$ kg makes contact with the spherical membrane and Fig.~\ref{fig:ballDemoPlot}(d) shows its equilibrium state.
Assuming the equilibrium state corresponds to $t = 0$ s, the spherical membrane undergoes further inflation at an increasing rate of $ \dot{p} = 100$ Pa/s.
Fig.~\ref{fig:ballDemoPlot}(e) illustrates the relationship between inflation time and the volume enclosed by the spherical membrane.
Initially, the enclosed volume is minimal, and during the early phases of inflation, the changes in the enclosed volume remain non-significant.
Upon reaching the limit point of the internal pressure, a sudden and significant volumetric snap-through occurs, simultaneously actuating the ball upwards.
This is evidenced by a pronounced increase in volume as depicted in the figure.
Furthermore, when the ball detaches from the spherical membrane, there is a noticeable alteration in the rate of volume change.
Applying an electric potential difference across the thickness of the membrane leads to the snap-through occurring at an earlier stage. 
This earlier occurrence of snap-through is also accompanied by a more pronounced volumetric change.
As the electric potential difference increases, the snap-through occurs at earlier states, and greater enclosed volumes are also achieved.
Fig.~\ref{fig:ballDemoPlot}(f) illustrates the variation in the height of the ball as a function of time for increasingly applied potential differences.
Prior to reaching the limit pressure, the height of the ball experiences a gradual increment.
After the volumetric snap-through induced by inflation, the ball is actuated, exhibiting motion characterized by a parabolic trajectory.
The increase of the applied potential differences results in the ball attaining a more elevated position.
Consequently, the soft actuator enhanced by the electroelastic coupling effect can actuate the ball to ascend a greater height compared to the pure mechanical case.
In other words, integrating the electroelastic coupling effect enhances the spherical actuator's capabilities.

\section{Computational efficiency}
\label{sec:computeTime}

\begin{figure}[t!]
  \centering
  \includegraphics[width=0.8\textwidth]{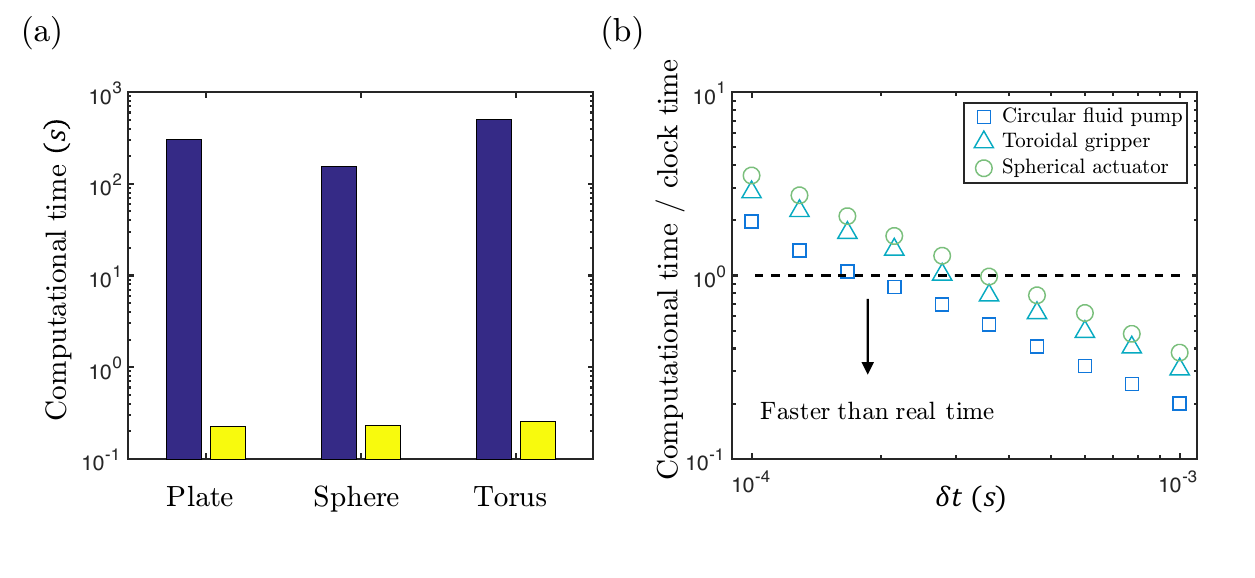}
1  \caption{Computational efficiency analysis of the proposed 1D DDG-based model (a) The comparison of computational time between the proposed method (shown in yellow) and a 3D FEM shell model (shown in blue) is conducted using hyperelastic benchmarks. (b) Computational time is plotted against wall-clock time as a function of the time step size for dynamic scenarios.}
  \label{fig:timePlot}
\end{figure}

Incorporating axisymmetry into the DDG-based model offers a significant advantage by simplifying intricate 3D geometries into 1D discrete models.
This axisymmetric property effectively simplifies the computational representation, leading to a reduction in both computational complexity and resource demands, while still maintaining the accuracy of the numerical solution.
By leveraging the simplified geometric and topological structure of the 1D model, the approach facilitates the development of more efficient numerical algorithms and strategies.
This optimization further improves computational efficiency and leads to reduced computational costs, enabling the simulation of complex 3D problems in real-time.
Figure~\ref{fig:timePlot}(a) compares the computational time of the proposed method (shown in yellow) with that of a three-dimensional finite element shell model (shown in blue).
Both methods address benchmark problems illustrated in Section~\ref{sec:hyperModelvalidation}, aiming to achieve comparable accuracy levels.
To simulate the nonlinear mechanical response of a hyperelastic membrane, the FEM-based 3D framework proposed in~\cite{liu2023computational} would take multiple minutes, while the newly-introduced 1D discrete model can predict the result within one second.
The simulations are performed on a single thread of an Intel Core i7 - 6600U Processor @ 3.4 GHz on a desktop processor.
The proposed method significantly reduces computational time compared to the conventional 3D finite element method.
On the other side, our model can achieve faster than real-time simulation for dynamic analysis.
Figure~\ref{fig:timePlot}(b) displays the computational time for three robotic application simulations plotted against wall-clock time, indicating that the proposed method can simulate complex engineering problems in real time as long as a suitable time step size is selected.
This capability is a prerequisite for both optimal design and online control of soft robots and actuators, e.g., iterating over a wide variety of parameters to find a robot design or actuation strategy.

\section{Concluding remarks}
\label{sec:conclusions}
This manuscript proposed an innovative DDG-based approach for the simulation and analysis of axisymmetric electroelastic membranes subjected to both inflation and electrical load.
It proficiently addresses not only contact but also dynamic challenges.
First, the method is validated using hyperelastic benchmarks, including a pressurized circular plate, a spherical balloon, and a toroidal membrane.
The numerical results are compared with both analytical solutions and alternative methods, revealing high agreement that proves the effectiveness of the proposed method.
The investigation of electroelastic spherical and toroidal membranes is additionally employed to validate the proposed method in addressing electro-mechanical coupling problems.
It becomes evident that not only does the internal pressure exhibit a limiting point, but the electrical load also manifests a maximum threshold.
Consequently, a snap-through phenomenon can be induced by the electrical load.
This observation substantiates that the inflated electroelastic membranes can be conceptualized as soft actuators activated by the application of an electrical load.
Three robotic applications utilizing the characteristics of the inflated electroelastic membrane are introduced.
The complete pumping mechanism of a soft circular fluid pump is simulated.
It capitalizes on the reversible snap-through feature of the electroelastic membrane to achieve significant volume changes.
A soft toroidal gripper is conceptualized to grasp objects through volumetric snap-through triggered by either mechanical or electrical loads.
The toroidal gripper can also enter the void spaces of an object and inflate from within to make contact with the object, ensuring secure extraction.
Applying an electrical load will advance the onset of the snap-through phenomenon while also increasing the contact force.
Another presented application features a soft actuator utilizing an electroelastic spherical membrane, which is capable of moving a rigid ball through inflation.
The volumetric snap-through of the inflated spherical membrane results in upward propulsion of the ball, and the application of an electrical load facilitates the elevation of the ball to a higher altitude.

\section*{Acknowledgments}

M.L. acknowledges the startup funding from the University of Birmingham. K.J.H. acknowledges the financial support from Nanyang Technological University, Singapore (Grant M4082428), and the Ministry of Education, Singapore, under its Academic Research Fund Tier 2 (T2EP50122-0005).  X.H. acknowledges the financial support from the startup funding from University of Michigan, Ann Arbor. W.H. acknowledges the financial support from the startup funding from Newcastle University. We are grateful to Huichan Zhao for the useful discussions.

\section*{Conflict of interest}
The authors declare that they have no conflict of interest.

\section*{Availability of data}
The datasets generated during the current study are available from the corresponding author upon reasonable request.

\section*{Availability of Code}
The code generated during the current study is available from the corresponding author on reasonable request.

\bibliographystyle{vancouver}
\bibliography{template}
\end{document}